\documentclass[preprint,superscriptaddress]{revtex4}
\usepackage{graphicx}
\usepackage{array}
\usepackage{amsfonts}
\usepackage{amssymb,amsmath,multirow,rotate}
\usepackage{mathrsfs}
\usepackage{booktabs}
\usepackage{threeparttable}
\usepackage{multirow}
\usepackage{epsfig}
\usepackage{threeparttable}
\usepackage{chngpage}
\usepackage{latexsym}
\usepackage{dcolumn}
\usepackage{graphics,graphicx,bm,epic,eepic,float}
\usepackage{times}
\usepackage{verbatim}
\usepackage{subfigure}
\usepackage{color}
\usepackage{tabularx}
\newcolumntype{Z}{>{\centering\arraybackslash}X} 
\definecolor{red}{rgb}{1,0,0}

\definecolor{blue}{rgb}{0,0,1}

\begin{document}

\title{Collective dynamics of identical phase oscillators with high-order coupling}

\date{\today}
\author{Can Xu}
\affiliation{Department of Physics and the Beijing-Hong Kong-Singapore Joint Center for Nonlinear and Complex Systems (Beijing), Beijing Normal University, Beijing 100875, China}

\author{Hairong Xiang}
\affiliation{Department of Physics and the Beijing-Hong Kong-Singapore Joint Center for Nonlinear and Complex Systems (Beijing), Beijing Normal University, Beijing 100875, China}

\author{Jian Gao}
\affiliation{Department of Physics and the Beijing-Hong Kong-Singapore Joint Center for Nonlinear and Complex Systems (Beijing), Beijing Normal University, Beijing 100875, China}

\author{Zhigang Zheng}\email{zgzheng@hqu.edu.cn}
\affiliation{College of Information Science and Engineering, Huaqiao University, Xiamen 361021, China}

\begin{abstract}
{\bf In this paper, we propose a framework to investigate the collective dynamics in ensembles of globally coupled phase oscillators when higher-order modes dominate the coupling. The spatiotemporal properties of the attractors in various regions of parameter space are analyzed. Furthermore, a detailed linear stability analysis proves that the stationary symmetric distribution is only neutrally stable in the marginal regime which stems from the generalized time-reversal symmetry. Moreover, the critical parameters of the transition among various regimes are determined analytically by both the Ott-Antonsen method and linear stability analysis, the transient dynamics are further revealed in terms of the characteristic curves method. Finally, for the more general initial condition the symmetric dynamics could be reduced to a rigorous three-dimensional manifold which shows that the neutrally stable chaos could also occur in this model for particular parameter. Our theoretical analysis and numerical results are consistent with each other, which can help us understand the dynamical properties in general system with higher-order harmonics couplings.}
\end{abstract}

\maketitle
Large system of coupled oscillators occur in a wide variety of situation throughout the nature which has attracted much attention from the scientific community during the last decades~\cite{pikovsky2001synchronization}. Examples are including the flashing of fireflies~\cite{buck1988synchronous}, electrochemical and spin-toque oscillators~\cite{georges2008impact,kiss2002emerging}, pedestrians on footbridges~\cite{eckhardt2007modeling}, applauding person in a large audience~\cite{neda2000physics} and others. Understanding the cooperative dynamical properties of such system is therefore of considerable theoretical and experiment interest. Indeed, in the weak interaction limit, the dynamics of limit-cycle oscillators could be effectively described in terms of their phase variable $\theta$, while the most famous case is the Kuramoto model~\cite{kuramoto1984chemical} which stands out as the classical paradigm for studying the spontaneous emergence collective synchronization in such system~\cite{strogatz2000kuramoto}. The form of phase equation obeys
\begin{equation}\label{equ:in01}
\dot{\theta}_{i}=\omega_{i}+\sum_{j=1}^{N}\Gamma(\theta_{j}-\theta_{i})\,,
\end{equation}
where $\theta_{i}$ denotes the phase of the $i$th oscillators, $\omega_{i}$ is its natural frequency. $\Gamma(\theta)$ is $2\pi$-periodic function representing the interaction between units. The simple choice of $\Gamma(\theta)=(K/N)\sin\theta$ leads to the classical Kuramoto model. Along the past decades the Kuramoto model with its generalizations have inspired and simulated a wealth of extensive studies because of both their simplicity for mathematical treatment
and their relevance to practice~\cite{acebron2005kuramoto,arenas2008synchronization,rodrigues2015kuramoto}. In particular, when $\Gamma(\theta)$ includes higher harmonics the system exhibits nontrivial dynamical features which were reported in the recent works~\cite{skardal2011cluster,komarov2013multiplicity,komarov2014kuramoto,li2014transition}.

In many realistic systems, the higher harmonic term (especial the second) always plays an essential role in the interaction and even dominates the coupling function, such as the Huygens pendulum system~\cite{czolczynshi2013synchronization}, the neuronal oscillators and genetic networks system~\cite{zhang2009synchronization}, the globally coupled photochemical oscillators system~\cite{kiss2005predictiog}, etc. In contrast to the pervious discussions of Kuramoto model containing higher order coupling. In the present work, we study phase equations of the following form
\begin{equation}\label{equ:in02}
\dot{\theta}_{i}=\omega-\lambda\sin 2\theta_{i}+\dfrac{\sigma}{N}\sum_{j=1}^{N}\sin 2\theta_{j},\; i=1,\cdots, N,
\end{equation}
where $\omega$ is the frequency of identical oscillators, $\lambda$ and $\sigma$ are the coupling strength respectively. There are various motivations for the study of this model, for examples, in the Josephson junction arrays~\cite{goldobin2011josephson,goldobin2013phase} the dynamics of a single element is determinated by a double well potential and therefore the strong effects caused by the second harmonics is important. Another example is the star-like model with a central element while the case of single harmonic term was considered in the paper~\cite{gomez2011explosive,zou2014basin,xu2015explosive,coutinho2013kuramoto,kazanovich2003synchronization,burylko2011bifurcations,kazanovich2013competition,vlasov2015explosive,vlasov2015star}. Similar to mean-field coupling in the Kuramoto model, the interaction term in Eq.(2) is a driving force that does not depends on the phase of driven oscillators and is equal on every oscillator.

In this paper, we present a complete framework to investigate the collective dynamics of globally coupled phase oscillator when higher-order modes dominate the coupling function. It includes several aspects, which together presents a global picture for the understanding of the dynamical properties in this system. First, we use the Ott-Antonsen method~\cite{ott2008low} to obtain the low-dimensional description of symmetric dynamics, where various states are illustrated schematically in the phase-diagram (Fig.1), they are double-center, single-center, center-synchrony coexistence, and synchrony regime respectively. Furthermore, a detailed linear stability analysis which is based on the self-consistent method is implemented and both the boundary curves and eigenvalues of steady states are obtained analytically which are consistent with the Ott-Antonsen method. Additionally, it has been proved that the linearized operator for the stationary symmetric distribution has infinite many pure imaginary eigenvalues which implies the stationary symmetric distribution is neutrally stable to perturbation in all the directions. Second, the two-cluster synchrony state is determined which is initial values dependent and the general transient solutions of the distribution are calculated in term of the characteristics method. Finally, the general initial conditions for the phase oscillators which lie off the poisson submanifold are considered where the symmetric dynamics are governed by the M\"{o}bius transformation which are rigorous three dimensional. And therefore one can expect the chaotic behavior of both the symmetric dynamics $r_{2}$ and the degree of asymmetry $r_{1}$ occur in the marginal regime. Extensive numerical simulations have been carried out to verify our theoretical analysis. In the following we report our main results both theoretically and numerically.

\bigskip
\noindent{\large\bf Results}\\
\noindent{\bf Symmetric dynamics.} We start by considering the high-order coupled phase oscillators model (2), without loss of generality the range of the coupling strength is restricted to $\lambda>0$ and $-\infty<\sigma<+\infty$ throughout the paper. The most important characteristic of the current model is the introduction of higher harmonics in the coupling function, and hence the generalized order parameter is needed to symbol the collective behavior of the system~\cite{skardal2011cluster} which yields
\begin{equation}\label{equ:re03}
r_{n}=R_{n}e^{i\Theta_{n}}=\dfrac{1}{N}\sum_{j=1}^{N}e^{in\theta_{j}}\,,
\end{equation}
for $n\in$ integer, where $R_{n}$ is the magnitude of the complex, $n$-th order parameter, and $\Theta_{n}$ is its phase. As named in~\cite{skardal2011cluster}, the amplitude $R_{2}$ measures the level of cluster synchrony while $R_{1}$ measures the degree of asymmetry in clustering. Eq.~(\ref{equ:in02}) can be rewritten in terms of $r_{2}$ as
\begin{equation}\label{equ:re04}
\dot{\theta}_{i}=\omega-\dfrac{\lambda}{2i}(e^{2i\theta_{i}}-e^{-2i\theta_{i}})+\sigma\cdot\mathrm{Im}(r_{2})\,,
\end{equation}
$\mathrm{Im}$ represents the imaginary pant. In the thermodynamic limit $N\rightarrow\infty$, Eq.~(\ref{equ:re04}) is equivalent to the continuity equation as a consequence of the conservation of the number of oscillators, i.e,
\begin{equation}\label{equ:re05}
\dfrac{\partial\rho}{\partial t}+\dfrac{\partial}{\partial \theta}\left\{\rho\cdot \left[\omega-\dfrac{\lambda}{2i}(e^{2i\theta}-e^{-2i\theta})+\sigma\cdot\mathrm{Im}(r_{2})\right]\right\}=0\,.
\end{equation}
Here $\rho(\theta,t)d\theta$ gives the fraction of oscillators which lie between $\theta$ and $\theta+d\theta$ at time $t$ with the appropriate normalization condition $\int_{0}^{2\pi}\rho(\theta,t)d\theta=1$, as a result the continuity limit of the generalized order parameter takes the form
\begin{equation}\label{equ:re06}
r_{n}(t)=\int_{0}^{2\pi}e^{in\theta}\rho(\theta,t)d\theta.
\end{equation}
Additionally, considering the $2\pi$-periodic of $\theta$ in the distribution function $\rho(\theta,t)$ which allows a Fourier expansion and can be written as
\begin{equation}\label{equ:re07}
\rho(\theta,t)=\dfrac{1}{2\pi}\sum_{n=-\infty}^{\infty}r_{n}(t)e^{-in\theta}=\rho_{s}+\rho_{a}\,.
\end{equation}
It is obvious that the $n$-th Fourier coefficient of $\rho(\theta,t)$ is just the $n$-th order parameter Eq.~(\ref{equ:re06}). Here $\rho_{s}$ is the sum where $n$ is even and is symmetric in the sense that $\rho_{s}(\theta+\pi,t)=\rho_{s}(\theta,t)$, and $\rho_{a}$ is the sum where $n$ is odd which is antisymmetric with respect to the translation by $\pi$, $\rho_{a}(\theta+\pi,t)=-\rho_{a}(\theta,t)$. Then, substituting the expansion Eq.~(\ref{equ:re07}) into the continuity Eq.~(\ref{equ:re05}) we obtain a set of two equations
\begin{equation}\label{equ:re08}
\dot{r}_{2n+1}=i(2n+1)\left[\dfrac{i\lambda}{2}\cdot r_{2n+3}-\dfrac{i\lambda}{2}r_{2n-1}+(\omega+\sigma\mathrm{Im}(r_{2}))\cdot r_{2n+1}\right],
\end{equation}
for the odd Fourier coefficient and
\begin{equation}\label{equ:re09}
\dot{r}_{2n}=i(2n)\left[\dfrac{i\lambda}{2}\cdot r_{2n+2}-\dfrac{i\lambda}{2}r_{2n-2}+(\omega+\sigma\mathrm{Im}(r_{2}))\cdot r_{2n}\right],
\end{equation}
for the even Fourier coefficient. Eq.~(\ref{equ:re08}) together with Eq.~(\ref{equ:re09}) provide two set of infinite many coupled equations which evolve independently. For instance, the motion of $r_{2}$ is not only dependent on itself but also governed by $r_{4}$ and $r_{0}$. However, observing the specific form of Eq.~(\ref{equ:re08}) and Eq.~(\ref{equ:re09}) one notices that Eq.~(\ref{equ:re08}) has a trivial invariant manifold solution
\begin{equation}\label{equ:re10}
r_{n}\equiv 0,\qquad n\in odd,
\end{equation}
and Eq.~(\ref{equ:re09}) has a non-trivial invariant manifold solution
\begin{equation}\label{equ:re11}
r_{n}\equiv r_{2}^{n},\qquad n\in even,
\end{equation}
which is indeed the Ott-Antonsen ansatz~\cite{ott2008low,ott2009long}. The Ott-Antonsen method yields a special solution for system provided $r_{2}$ evolves according to a single ordinary differential equation
\begin{equation}\label{equ:re12}
\dot{r}_{2}=-\lambda r_{2}^{2}+\lambda+2i(\omega+\sigma\mathrm{Im}(r_{2}))\cdot r_{2}\,,
\end{equation}
solution of this kind turn out to form a two-dimensional invariant manifold which is the set of Poisson kernels
\begin{equation}\label{equ:re13}
\rho_{s}(2\theta,t)=\dfrac{1}{2\pi}\dfrac{1-R_{2}^{2}}{1+R_{2}^{2}-2R_{2}\cos(\Theta_{2}-2\theta)}\,.
\end{equation}
One central issue in the study is to identify all the possible collective states both steady or nonstationary and reveal various bifurcations as the change of the coupling parameter $\lambda$ and $\sigma$ when the initial symmetric distribution has the form of Possion kernels.

The Riccati equation  Eq.~(\ref{equ:re12}) describes the collective symmetric dynamics of Eq.~(\ref{equ:in02}) in terms of the second order parameter $r_{2}$, and it can be straightforwardly rewritten in cartesian coordinates $r_{2}=x+iy$ as
\begin{eqnarray}
\label{equ:re14}
\dot{x}&=&-2(\omega+\sigma\cdot y-\lambda\, y)y+\lambda(1-x^{2}-y^{2})\,,\\
\label{equ:re15}
\dot{y}&=&2(\omega+\sigma\cdot y-\lambda\, y)x\,.
\end{eqnarray}
In the phase space of the second order parameter, the natural boundary is $x^{2}+y^{2}=1$ and a fixed point is determined by the intersection of nullclines $\dot{x}=0$ and $\dot{y}=0$ within the boundary. One recalls that the first point is $y_{01}=\omega/(\lambda-\sigma)$ and $x_{01}^{2}+y_{01}^{2}\equiv 1$ which is on the unit circle and defined as the synchrony state, the existence condition for the synchrony state is $|\lambda-\sigma|\geq\omega$. Additionally, the linearization Jacobian matrix of the synchrony state has two eigenvalues $\delta_{1}=-2\lambda x_{01}$ and $\delta_{2}=-2(\lambda-\sigma)x_{01}$. The same strategy such as the existence and the stability conditions can be actually adopted for the second fixed point $x_{02}=0$, $y_{02}=(\omega\pm\sqrt{\omega^{2}-\lambda(\lambda-2\sigma)}\,)/(\lambda-2\sigma)$, which is in the $y$ axis and termed as the splay state~\cite{watanabe1994constants}. The Jacobian matrix of the second point has two pure imaginary eigenvalues which shows that the splay state is neutral stable to perturbation and is a center in the Ott-Antonsen manifold.

\begin{figure}
  \includegraphics[width=1.00\linewidth,height=0.49\linewidth]{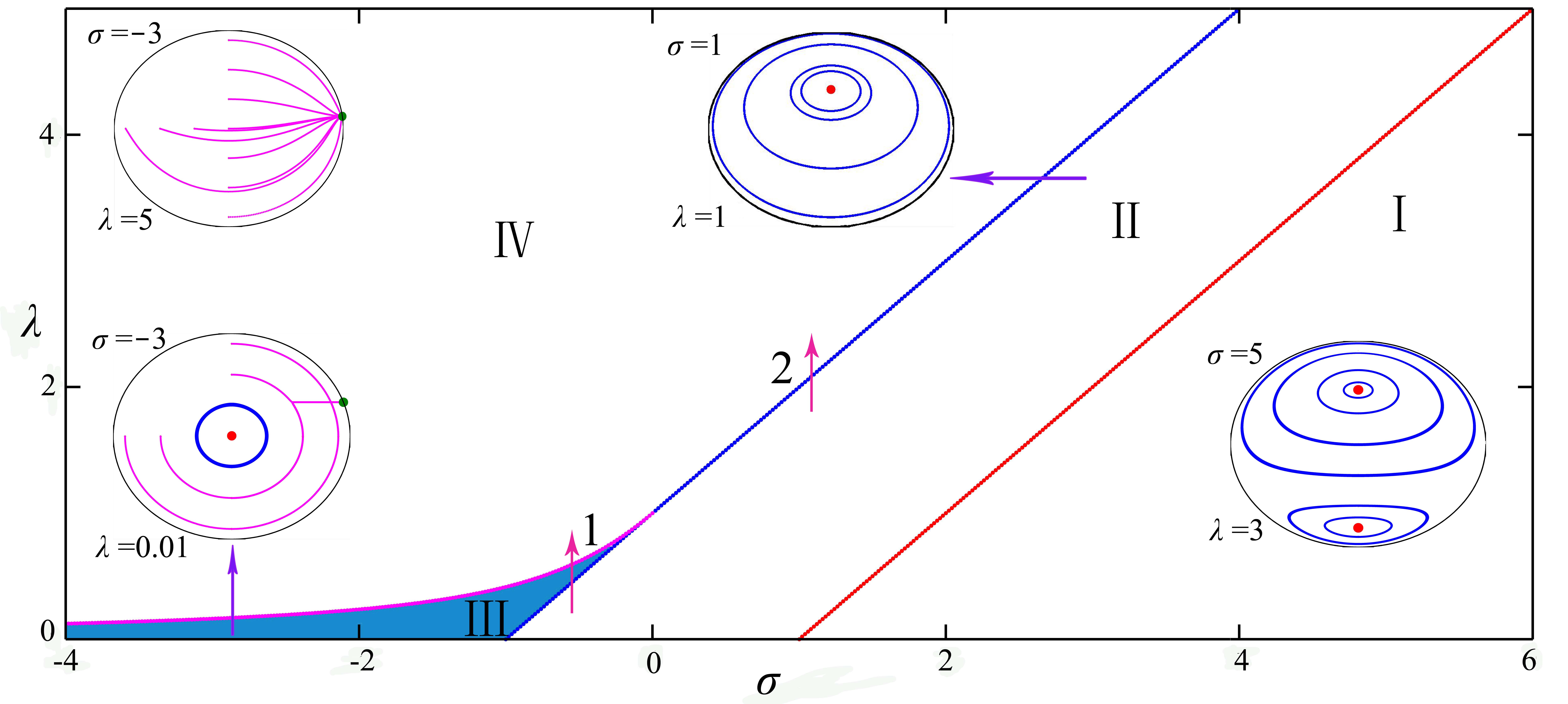}\\
 \caption{Phase diagram of the symmetric dynamics in the parameter space $\sigma$ and $\lambda$\,. \uppercase\expandafter{\romannumeral1} is double center regime, \uppercase\expandafter{\romannumeral2} is single center regime, \uppercase\expandafter{\romannumeral3} is center-synchrony coexistence regime, \uppercase\expandafter{\romannumeral4} is synchrony regime, respectively. The boundary curves are $\lambda=\sigma-\omega$ from \uppercase\expandafter{\romannumeral1} to \uppercase\expandafter{\romannumeral2}\,, $\lambda=\sigma+\omega$ from \uppercase\expandafter{\romannumeral2} to \uppercase\expandafter{\romannumeral3} or \uppercase\expandafter{\romannumeral4}\,, $\lambda=\sigma+\sqrt{\omega^{2}+\sigma^{2}}$ from \uppercase\expandafter{\romannumeral3} to \uppercase\expandafter{\romannumeral4}, in the numerical simulations we set $\omega=1$.
 }\label{figure1}
\end{figure}

Fig.~\ref{figure1} summarizes the results of our analysis of Eq.~(\ref{equ:re14}) and Eq.~(\ref{equ:re15}). We find that there are four types of regime in the phase diagram, they are the double-center region \uppercase\expandafter{\romannumeral1}, single-center region \uppercase\expandafter{\romannumeral2}, center-synchrony coexistence region \uppercase\expandafter{\romannumeral3} and the synchrony region \uppercase\expandafter{\romannumeral4} respectively. In addition, various bifurcations and transitions among the states are illustrated when the parameters pass through the boundary curves. There are two kinds of route to synchrony, for example, when we start from \uppercase\expandafter{\romannumeral2} to \uppercase\expandafter{\romannumeral4} (arrow 1 in the Fig.~\ref{figure1}), the transition to synchrony is a first-order phase transition with hysteresis because of the coexistence region \uppercase\expandafter{\romannumeral3}. however, when we turn to the second route (arrow 2 in the Fig.~\ref{figure1}) the transition is discontinuity which is absent of hysteresis.

The analysis above reveals the low-dimensional collective behavior of symmetric dynamics where the bifurcations and transition among various states are presented. However, all the results are within the framework of Ott-Antonsen invariant manifold including the linear perturbation of the fixed points. In the following, we conduct $\alpha$ through linear stability analysis of the symmetric dynamics in terms of the self-consistent theory~\cite{golomb1991clustering} where all the boundary curves could be obtained analytically in an alternative way. In particular, we show that when the perturbation is not within the invariant manifold the eigenvalues of the steady states keep the same form above and the splay state is neutral stable in all the directions.

A convenient way of solving for the symmetric dynamics is to make the change of variable $\varphi=2\theta$, which yields a new dynamical equation
\begin{equation}\label{equ:re16}
\dot{\varphi}_{i}=2\omega-2\lambda\sin\varphi_{i}+\dfrac{2\sigma}{N}\sum_{j=1}^{N}\sin\varphi_{j},\qquad i=1,\cdots,N.
\end{equation}
Eq.~(\ref{equ:re16}) has the form of linear Josephson junction arrays and it is obviously that the distribution of $\varphi$ is  equivalent to the $\rho_{s}$. The synchrony state is a spatially homogeneous fixed point of Eq.~(\ref{equ:re16}) defined as $\dot{\varphi}_{i}=0$, for all the $i$ and $\varphi_{1}=\varphi_{2}=\cdots=\varphi_{N}=\varphi_{0}$ which is the simplest attractor. Therefore, this phase leads to a solution of equation
\begin{equation}\label{equ:re17}
\omega-(\lambda-\sigma)\sin\varphi_{0}=0.
\end{equation}
In this case $\sin\varphi_{0}=\omega/(\lambda-\sigma)$, the Jacobian matrix of Eq.~(\ref{equ:re16}) is a circulant matrix~\cite{xu2015explosive,golomb1991clustering} which has two kinds of eigenvalues, the first one is
\begin{equation}\label{equ:re18}
\delta_{1}=-2(\lambda-\sigma)\cos\varphi_{0},
\end{equation}
that corresponds to a spatial homogeneous fluctuation and the other eigenvalues
\begin{equation}\label{equ:re19}
\delta_{2}=-2\lambda\cos\varphi_{0},
\end{equation}
which have $(N-1)$-fold degeneracy that correspond to inhomogeneous fluctuations. Incidentally, one notes that $x_{01}\equiv \cos\varphi_{0}$, $y_{01}\equiv\sin\varphi_{0}\,$, this result is consistent with a basic fact that the synchrony state is contained in the Ott-Antonsen manifold. Numerical experiment shows that the basin of attraction of the synchrony state in \uppercase\expandafter{\romannumeral4} is globally.

Another scenario is the stationary distribution where the phase $\varphi$ is smoothly distributed over $[0,\,2\pi]$ and from Eq.~(\ref{equ:re16}) one observes that the general form of such a stationary distribution is
\begin{equation}\label{equ:re20}
\rho_{s}(\varphi)=\dfrac{C}{\bar{\omega}-2\lambda\sin\varphi},
\end{equation}
where $\bar{\omega}$ is the effective frequency which could be determined self-consistently from the equation
\begin{equation}\label{equ:re21}
\bar{\omega}=2\omega+2\sigma\int_{0}^{2\pi}d\varphi\,\dfrac{C\cdot \sin\varphi}{\bar{\omega}-2\lambda\sin\varphi},
\end{equation}
and $C$ is the normalization constant $C=\pm\sqrt{\bar{\omega}^{2}-4\lambda^{2}}/2\pi$ when $\bar{\omega}>0$, $C$ is a positive one and vice verse. Substituting the expression of $C$ into Eq.~(\ref{equ:re21}) and after some calculation. We obtain the solution of $\bar{\omega}$
\begin{equation}\label{equ:re22}
\bar{\omega}=\dfrac{2\left[\omega(\lambda-\sigma)\pm\sqrt{\sigma^{2}(\omega^{2}-\lambda^{2}+2\lambda\sigma)}\right]}{\lambda-2\sigma}.
\end{equation}
It should be emphasized that the stationary distribution $\rho_{s}(\varphi)$ corresponds to the splay state in the invariant manifold above, because the distribution Eq.~(\ref{equ:re20}) has the form of Possion kernels and
\begin{equation}\label{equ:re23}
\langle\sin\varphi\rangle=\int_{0}^{2\pi}\dfrac{C\cdot\sin\varphi}{\bar{\omega}-2\lambda\sin\varphi}\,d\varphi=y_{02},
\end{equation}
and
\begin{equation}\label{equ:re24}
\langle\cos\varphi\rangle=\int_{0}^{2\pi}\dfrac{C\cdot\cos\varphi}{\bar{\omega}-2\lambda\sin\varphi}\,d\varphi=0\equiv x_{02}.
\end{equation}
The perturbation of the continuum equation for the stationary symmetric distribution $\rho_{s}(\varphi)$ is
\begin{equation}\label{equ:re24'}
\begin{split}
\dfrac{\partial}{\partial t}(\delta \rho)=&\hat{L}\,\delta\rho(\varphi,t)\\
=&-\dfrac{\partial}{\partial\varphi}\left[(\bar{\omega}-2\lambda\sin\varphi)\,\delta\rho(\varphi,t)\right]+\dfrac{\partial\rho_{s}}{\partial\varphi}\int_{0}^{2\pi}d\varphi'\,2\sigma\sin\varphi'\,\delta\rho(\varphi',t),
\end{split}
\end{equation}
and the eigenvalue equation Eq.~(\ref{equ:re24'}) is convenient to treat in the function space
\begin{equation}\label{equ:re24''}
\dfrac{\delta\rho(\varphi,t)}{\rho_{s}(\varphi)}=\sum_{n=-\infty}^{\infty}\,a_{n}(t)e^{2\pi in G(\varphi)},
\end{equation}
where $G(\varphi)$ is the basis function
\begin{equation}\label{equ:re24'''}
G(\varphi)=\int_{0}^{\varphi}\,\rho_{s}(\varphi')\,d\varphi',
\end{equation}
and $a_{n}(t)$ are the expansion coefficients. From the stability analysis of the stationary distribution (all the details are included in the supplementary material) we find that in the regime \uppercase\expandafter{\romannumeral1}, \uppercase\expandafter{\romannumeral2} and \uppercase\expandafter{\romannumeral3}, all the infinite many eigenvalues of the operator $L$ are pure imaginary which implies that the stationary distribution is neutrally stable in all the directions and it is not only confined to Ott-Antonsen manifold. The regime where the stationary distribution is marginally stable is termed as the marginal regime and in this regime there is no particular attractor that the system converges to~\cite{golomb1991clustering}, as a result, the highly non-generic property is associated with the time-reversal symmetry that the Eq.~(\ref{equ:re16}) exhibits. When we start in the Ott-Antonsen manifold the trajectory of $r_{2}$ is two-dimensional closed   periodic curve (the insertion of Fig.~\ref{figure1}), however, the situation differs substantially when we start with a more general initial condition in the marginal regime as we see in the following part.

\begin{figure}
  \includegraphics[width=1.00\linewidth,height=0.42\linewidth]{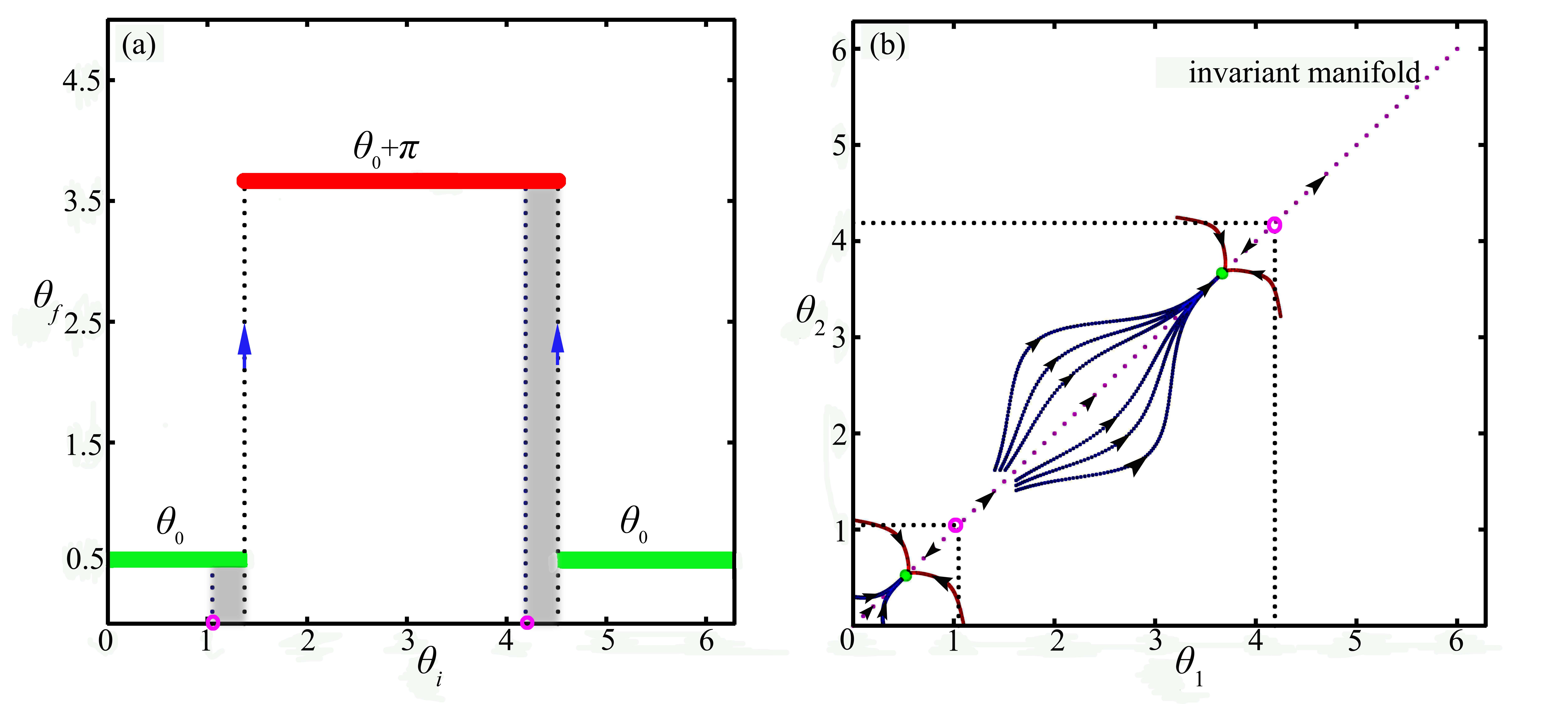}\\
 \caption{(a) The initial phase $\theta_{i}$ vs the final phase $\theta_{f} $ when we choose the initial phase distribution $\rho(\theta,t_{0})=(1+\cos \theta)/2\pi$ and the number of oscillator is $N=10000$ in the simulation, when the initial phase is in the gray area these oscillators will bypassing the saddle points (the pink hollow circle in the horizontal axis) which leads to $\varepsilon\neq0$.
 (b) The diagram of the Arnold diffusion mechanism in the same parameter with $N=3$ in terms of phase space of $\theta_{1}$ and $\theta_{2}$, for some particular trajectory such as the red line the oscillator will bypassing the saddle point in the invariant manifold.}\label{figure2}
\end{figure}

\bigskip
\noindent{\bf The steady and transient dynamics.} The analysis above investigate the symmetric dynamics by using two kinds of ways, one recalls that when the parameters are in regime \uppercase\expandafter{\romannumeral4}, the symmetric dynamics converge to steady state, and the two cluster synchrony states emerge $\sin 2\theta_{0}=\omega/(\lambda-\sigma)$, $\cos 2\theta_{0}>0\,$ accordingly. Thus, the complete steady state distribution of oscillators is
\begin{equation}\label{equ:re25}
\rho_{0}(\theta)=(\frac{1}{2}+c)\delta(\theta-\theta_{0})+(\frac{1}{2}-c)\delta(\theta-\theta_{0}-\pi),\qquad |c|<\frac{1}{2},
\end{equation}
 at this time $\rho_{0}(\theta)$ is comprised of two delta functions denoting two clusters of oscillators at $\theta_{0}=0.5\arcsin\omega/(\lambda-\sigma)$ and $\theta_{0}+\pi$. Hence, the phase oscillators settle to one of the two stable equilibria while the unstable equilibria $\pi/2-\theta_{0}$ and $3\pi/2-\theta_{0}$ serve as the boundaries for the basin of attraction. Therefore the degree of asymmetry $|r_{1}|$ is
\begin{equation}\label{equ:re26}
|r_{1}|=|2c(\cos\theta_{0}+i\sin\theta_{0})|=2c\,,
\end{equation}
which depends on the free parameter $c$ and could be determined approximatively from initial condition, note that $1/2-c$ is just the fraction of oscillators in the locked phase $\theta_{0}+\pi$, namely
\begin{equation}\label{equ:re27}
c=\dfrac{1}{2}+\int_{\frac{\pi}{2}-\theta_{0}}^{\frac{3\pi}{2}-\theta_{0}}\rho(\theta,t_{0})\,d\theta\,+\varepsilon,
\end{equation}
$\rho(\theta,t_{0})$ is the initial density, $\varepsilon$ is the error caused by the Arnold diffusion. Theoretically, when the initial phase is in the one dimensional invariant manifold $\theta_{1}=\theta_{2}=\cdots=\theta_{N}$ the evolution of phase $\theta(t)$ for all the oscillators can never pass through the two saddle points $\pi/2-\theta_{0}$ and $3\pi/2-\theta_{0}$ which means $\epsilon=0$. However, for the more general initial conditions some oscillators which are in the neighborhood of two unstable equilibrium states can bypassing the saddle points (Fig.~\ref{figure2}(b) illustrated schematically the mechanism in the low dimensional phase space where $N=3$) and therefore $\varepsilon$ is non-ignorable. Fig.~\ref{figure2}(a) presents the numerical simulation when we choose $\theta_{0}=\pi/6$ and $\rho(\theta,t_{0})=(1+\cos \theta)/2\pi$, it is clear that the initial phases in the gray area eventually bypassing the saddle points (the pink hollow circle in the horizontal axis) and this interesting phenomenon implies that those oscillators tend to choose a relatively near equilibrium state to settle in the high dimensional phase space while this is forbidden in the one dimensional invariant manifold.

\begin{figure}
  \includegraphics[width=1.00\linewidth,height=0.42\linewidth]{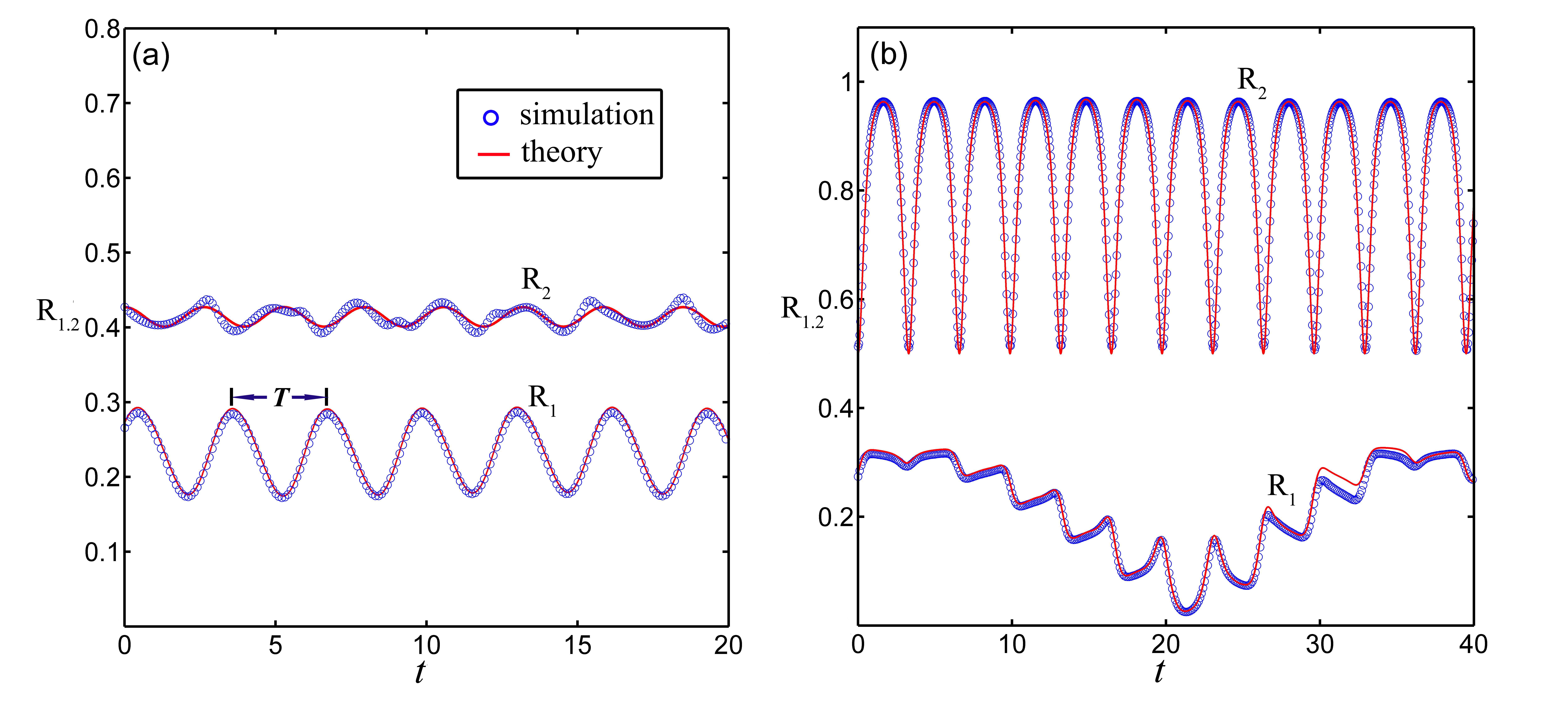}\\
 \caption{Transient dynamics of $R_{1}$ and $R_{2}$. (a) The amplitude of $R_{2}$ is small, $R_{1}$ oscillates regularly with a period $T$, $\sigma=1.0$, $\lambda=1.0$, $\mathrm{Re}(r_{20})=0$, and $\mathrm{Im}(r_{20})=0.427$. (b) The amplitude of $R_{2}$ is large, $R_{1}$ oscillates irregularly, $\sigma=1.0$, $\lambda=1.4$, $\mathrm{Re}(r_{20})=0$, and $\mathrm{Im}(r_{20})=-0.5$. In the numerical simulation we choose the initial phase distribution $\rho(\theta,t_{0})=\rho_{s}(2\theta,t_{0})(1+ 0.5\cos(\theta))$ and $N=10000$, the line is calculated by the characteristic theory and the circle is the numerical  simulation.
 }\label{figure3}
\end{figure}

When the symmetric dynamics $r_{2}$ gets to steady state, then the $|r_{1}|$ dynamics reaches steady state quickly. However, in a large marginal regime of the phase diagram Fig.~\ref{figure1}, the dynamics of $r_{2}$ can never converge to an attractor, the motion of $r_{2}$ is time-dependent. To capture the dynamics of $r_{1}$ we can solve the partial differential equation (PDE) Eq.~(\ref{equ:re05})
\begin{equation}\label{equ:re29}
\dfrac{\partial\rho}{\partial t}+(\omega-\lambda\sin 2\theta+\sigma\cdot\mathrm{Im}(r_{2}))\dfrac{\partial\rho}{\partial\theta}=\lambda\cos 2\theta\cdot\rho\,,
\end{equation}
in terms of the characteristics method, when we start in the characteristic curve $\theta(t,t_{0})$, the distribution function will be $\rho(\theta,t)\equiv\rho(\theta(t,t_{0}),t)$ and the PDE becomes the ODEs, the characteristic equations are
\begin{eqnarray}
\label{equ:re30}
&\dfrac{d\rho}{dt}=&2\lambda\cos 2\theta\cdot \rho\,,\\
\label{equ:re31}
&\dot{\theta}=&\omega-\lambda\sin 2\theta+\sigma\cdot\mathrm{Im}(r_{2}),
\end{eqnarray}
when the symmetric dynamics are in the Ott-Antonsen invariant manifold, the motion of $r_{2}$ is governed by the equation Eq.~(\ref{equ:re12}). Generally, the Eq.~(\ref{equ:re30}) and Eq.~(\ref{equ:re31}) are difficult to solve analytically while for some particular situation such as the orbit of $r_{2}$ is near to the center point, the amplitude of $r_{2}$ is small enough that $\mathrm{Im}(r_{2})$ approximates a constant, then the expression for the characteristic curves $\theta(t,t_{0})$ starting at the initial phase $\theta_{0}$ yields
\begin{equation}\label{equ:re32}
\theta(t,t_{0})=\arctan\left\{\dfrac{\lambda+\sqrt{(\omega+\sigma\mathrm{Im}(r_{2}))^{2}-\lambda^{2}}\,\cdot \tan[\sqrt{(\omega+\sigma\mathrm{Im}(r_{2}))^{2}-\lambda^{2}\,}(\theta_{0}+t)]}{\omega+\sigma\cdot\mathrm{Im}(r_{2})}\right\}
\end{equation}
and the distribution along the characteristic equations is
\begin{equation}\label{equ:re33}
\begin{split}
\rho(t)=&\rho_{0}\exp\left[\int_{t_{0}}^{t}2\lambda\cos 2\theta(t',t_{0})\,dt'\right]\\
=&\rho_{0}\exp\left[\int_{\theta_{0}}^{\theta(t)}2\lambda\cos 2\theta(t',t_{0})\cdot\dfrac{dt'}{d\theta}\cdot d\theta\right]\\
=&\dfrac{\rho_{0}(\omega-\lambda\sin 2\theta_{0}+\sigma\cdot \mathrm{Im}(r_{20}))}{\omega-\lambda\sin 2\theta(t,t_{0})+\sigma\cdot\mathrm{Im}(r_{2})}\,,
\end{split}
\end{equation}
where $\rho_{0}$ is the initial value of the distribution function,
$r_{20}$ is the initial value of $r_{2}$. Theoretically, the general form of density function $\rho(\theta,t)$ could be determined by substituting the inverse solution $\theta_{0}(\theta(t))$ Eq.~(\ref{equ:re32}) into Eq.~(\ref{equ:re33}), and the generalized order parameter could be calculated through the integral Eq.~(\ref{equ:re06}) while the difficult is due to the multivalue of anti-trigonometric function Eq.~(\ref{equ:re32}). Hence it is convenient to calculate the first-order parameter $r_{1}$ via the integral
\begin{equation}\label{equ:re34}
r_{1}(t)=\int_{-\pi}^{\pi}\rho(t)e^{i\theta(t,t_{0})}\cdot\dfrac{\partial\theta(t,t_{0})}{\partial\theta_{0}}\cdot d\theta_{0}\,,
\end{equation}
From Fig.~\ref{figure3}(a) we find that $|r_{1}(t)|$ oscillates regularly with a period
\begin{equation}\label{equ:re35}
T=\dfrac{\pi}{\sqrt{(\omega+\sigma\cdot\mathrm{Im}(r_{2}))^{2}-\lambda^{2}}}\,.
\end{equation}
when the amplitude of $r_{2}$ is considerable large, the time dependent of $r_{1}$ is irregular Fig.~\ref{figure3}(b).

The discussion of the symmetric dynamics Eq.~(\ref{equ:re16}) exhibits a two-dimensional Ott-Antonsen manifold proving that the initial symmetric distribution takes the form of Possion kernels Eq.~(\ref{equ:re13}). In fact, the significant works point out that the governing equations for the form of Eq.~(\ref{equ:re16}) are generated by the action of the M\"{o}bius group~\cite{marvel2009identical,marvel2009invariant}
\begin{equation}\label{equ:re36}
e^{i\varphi_{j}(t)}=\dfrac{e^{i\psi}e^{i\phi_{j}}+\alpha}{1+\alpha^{*}\, e^{i\psi}\cdot e^{i\phi_{j}}}\,,
\end{equation}
$\alpha$ is a complex variable, $\psi$ real, and $\phi_{j}$ motion constant. The group action partition the $N$-dimensional state space into three-dimensional invariant manifold and the three parameters $\mathrm{Re}(\alpha),\,\mathrm{Im}(\alpha),\,\psi$ are governed by the following equation
\begin{eqnarray}
\label{equ:re37}
\dot{\mathrm{Re}}(\alpha)&=&-2(\omega+\sigma\cdot\mathrm{Im}(r_{2})-\lambda\cdot\mathrm{Im}(\alpha))\cdot\mathrm{Im}(\alpha)+\lambda(1-|\alpha|^{2}),\\
\label{equ:re38}
\dot{\mathrm{Im}}(\alpha)&=&2(\omega+\sigma\cdot\mathrm{Im}(r_{2})-\lambda\cdot\mathrm{Im}(\alpha))\cdot\mathrm{Re}(\alpha),\\
\label{equ:re39}
\dot{\psi}&=&2(\omega+\sigma\cdot\mathrm{Im}(r_{2})-\lambda\cdot\mathrm{Im}(\alpha)),
\end{eqnarray}
when the motion constant takes a general distribution, the parameter $r_{2}$ can be written in terms of $\alpha$ and $\psi$ as~\cite{marvel2009identical}
\begin{equation}\label{equ:re40}
r_{2}=\alpha+(|\alpha|^{2}-1)\sum_{n=1}^{\infty}(-1)^{n}c_{n}^{*}e^{in\psi}(\alpha^{*})^{n-1}\,,
\end{equation}
$c_{n}$ is the $n$-th Fourier expansion coefficient of the distribution of motion constants. For the simple case when the distribution is uniform on $[0,2\pi]\,$, $c_{n}\equiv 0$ for all the $n\,$, $r_{2}\equiv \alpha$, and $\alpha$ decouples from $\psi$, this implies that the three-dimensional states phase has a two-dimensional invariant submanifold  which is indeed the Ott-Antonsen manifold. However in the more typical case that $\alpha$ and $\psi$ are interdependent the three-dimensional equation can exhibit non-general dynamical behavior in the marginal regime.

\begin{figure}
  \includegraphics[width=1.00\linewidth,height=0.49\linewidth]{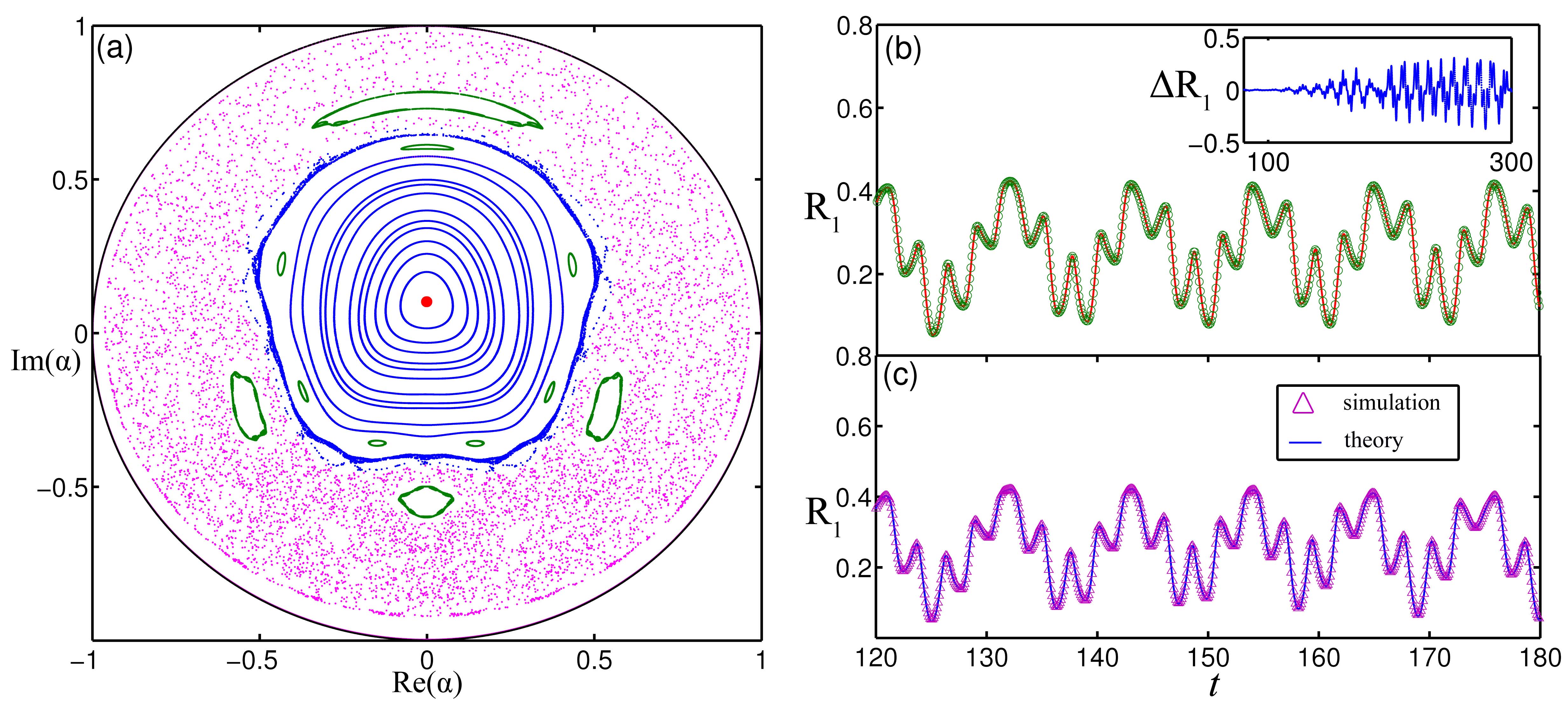}\\
 \caption{(a) Poincare section of Eq.~(\ref{equ:re37}) $\sim$ Eq.~(\ref{equ:re39}) at $\psi(\mod 2\pi)=0$, quasiperiodic trajectories appear as closed curves or island chains, periodic trajectories appear as fixed points or period-$p$ points of integer period, and chaotic trajectories fill the remaining region of the unit disk, where $\omega=1.0, \lambda=1.5, \sigma=2.0$. (b) the $R_{1}$ vs $t$ in the chaos regime with $\alpha(0)=0.5+i\,0.5$ and $\psi(0)=0.0$. (c) the $R_{1}$ vs $t$ with adjacent parameter value $\delta\alpha(0)=0.001$, $\delta\psi(0)=0$. The illustration is the difference of $R_{1}$ vs $t$ where we can see that the bias of the order parameter with neighboring parameters will be significant in the long time. In the simulation we choose $N=100000$, the line is determined in terms of the the characteristic theory and the circle and triangle are calculated by the numerical simulation.}\label{figure4}
\end{figure}

In Fig.~\ref{figure4}(a) we use Poincare section at $\,\psi(\mod 2\pi)=0\,$ to sort out the structure of state space, where the parameters are chosen in regime \uppercase\expandafter{\romannumeral2} of Fig.~\ref{figure1} ($\lambda=1.5$ and $\sigma=2.0$), and the motion constant has a distribution $(1+\sin\phi)/2\pi$. In the Poincare section, quasiperiodic trajectories appear as closed curves or island chains, periodic trajectories appear as fixed points or period-$p$ points of integer period, and chaotic trajectories fill the remaining regions of the unit disk, the picture of the phase portraits is reminiscent of Hamiltonian chaos and the appearance of the "quasi-Hamiltonian" properties reflecting the time reversibility symmetry under the transformation $t\rightarrow -t,\,\psi\rightarrow -\psi,\,x\rightarrow -x$, the system is invariant. When we choose the initial value in the chaos regime $\alpha(0)=0.5+i0.5$, $\psi(0)=0.0$, the three Lyapunov exponents are $\lambda_{1}=-0.266 \, $, $\lambda_{2}=0 \,$, $\lambda_{3}= 0.227\,$ respectively. As a result the order parameter $r_{2}$ and $r_{1}$ is also chaotic which is initial values sensitive. Fig.~\ref{figure4}(b) and (c) depicture the evolution of $R_{1}(t)$ with two adjacent parameter value where $\delta\alpha(0)=0.001$, $\delta\psi(0)=0$ and it is clear that the bias of order parameter with neighboring parameters (illustration in the Fig.~\ref{figure4}(b)) will be significant in the long time and both the characteristic curve and numerical simulation are consistent with each other well.

\bigskip
\noindent{\large\bf Discussion}
\\ \noindent
To summarize, we investigated the collective dynamics of globally coupled identical phase oscillator when second harmonics dominate the coupling and solutions can be decomposed into symmetric and antisymmetric part independently. Theoretically, Ott-Antonsen method, linear stability analysis, and characteristic method have been carried out to obtain insights. Together with the numerical simulations, our study presented the following main results. First, we obtain the low-dimensional description of the symmetric dynamics and various regimes are predicted in the phase diagram, including the double-center, the single-center, the center-synchrony coexistence, and the synchrony regime. Second, all the steady states and the boundary curves has been obtained analytically both in terms of the Ott-Antonsen anatz and linear stability analysis. Third, we proved that in the marginal regime the stationary symmetry distribution is only neutrally stable where all the infinitely many eigenvalues are pure imaginary. Finally, the characteristic method has been adopted to obtain the transient dynamics $r_{1}$  which evolves strongly depends on the initial values. Furthermore, for the general case of initial condition the symmetry dynamics are governed by the M\"{o}bius transformation and numerical experiment suggests that three-dimensional invariant manifolds contain neutrally stable chaos. This work provided a complete framework to deal with the high-order coupling phase oscillators model, and the obtained results will enhance our understandings of the dynamical properties of more harmonics coupling phase oscillator system.

\section*{Acknowledgements}
This work is partially supported by the NSFC grants Nos. 11075016,  11475022, 11135001, and the Scientific Research Funds of Huaqiao University.

\section*{Author contributions}
C.X., Y.T.S., J.G., T.Q, S.G.G and Z.G.Z. designed the research; C.X., Y.T.S. and T.Q performed numerical simulations and theoretical analysis; C.X., S.G.G and Z.G.Z. wrote the paper. All authors reviewed and approved the manuscript.

\section*{Additional information}
{\bf Competing financial interests:} The authors declare no competing financial interests.\\

Correspondence and requests for materials should be addressed to Z.G.Z. (zgzheng@bnu.edu.cn), or S.G.G. (guanshuguang@hotmail.com).

\clearpage

\end{document}